\documentclass[review]{elsarticle}
\usepackage{lineno,hyperref}
\modulolinenumbers[5]

\journal{Journal of \LaTeX\ Templates}

\begin{document}

\begin{frontmatter}

\title{The PANDA Barrel DIRC: From Design to Assembly}

\author[1]{G.~Schepers}

\author[1]{A.~Belias}

\author[1]{R.~Dzhygadlo}

\author[1]{A.~Gerhardt}

\author[1]{D.~Lehmann}

\author[1,2]{K.~Peters}

\author[1]{C.~Schwarz}

\author[1]{J.~Schwiening}

\author[1]{M.~Traxler}

\author[1,2]{Y.~Wolf}

\author[3]{L.~Schmitt}

\author[4]{M.~B\"{o}hm}

\author[4]{K.~Gumbert}

\author[4]{S.~Krauss}

\author[4]{A.~Lehmann}

\author[4]{D.~Miehling}

\author[5]{M.~Schmidt}

﻿﻿\author[6]{C.~Sfienti}

\author[7]{A.~Ali}

\address[1]{GSI Helmholtzzentrum f\"ur Schwerionenforschung GmbH , Darmstadt, Germany}

\address[2]{Goethe-University, Frankfurt, Germany}

﻿﻿\address[3]{FAIR, Facility for Antiproton and Ion Research in Europe, Darmstadt, Germany}

\address[4]{Friedrich Alexander-University of Erlangen-Nuremberg, Erlangen, Germany}

\address[5]{II. Physikalisches Institut, Justus Liebig-University of Giessen, Giessen, Germany}

\address[6]{Institut f\"{u}r Kernphysik, Johannes Gutenberg-University of Mainz, Mainz, Germany}

\address[7]{﻿﻿Helmholtz-Institut Mainz, Mainz, Germany}

\date{\vspace{-5ex}}


\address{GSI Helmholtzzentrum f\"ur Schwerionenforschung, Darmstadt, Germany}



\begin{abstract}
The Barrel DIRC counter will serve as the primary particle identification detector in the PANDA experiment, enabling high-precision hadron physics studies through antiproton-proton annihilations across a momentum range of 1.5 GeV/c to 15 GeV/c. It is designed to distinguish charged pions from kaons with a separation of at least 3~standard deviations up to 3.5 GeV/c within polar angles of 22$^\circ$ to 140$^\circ$. A lens focusing is used the first time in a DIRC detector. After a successful evaluation in particle beams, the key components, i.e. radiator bars and photon sensors, were purchased and tested.
\end{abstract}
\end{frontmatter}
\section{Design}
\begin{figure}[h]
  \centering
  \includegraphics[width=0.90\columnwidth]{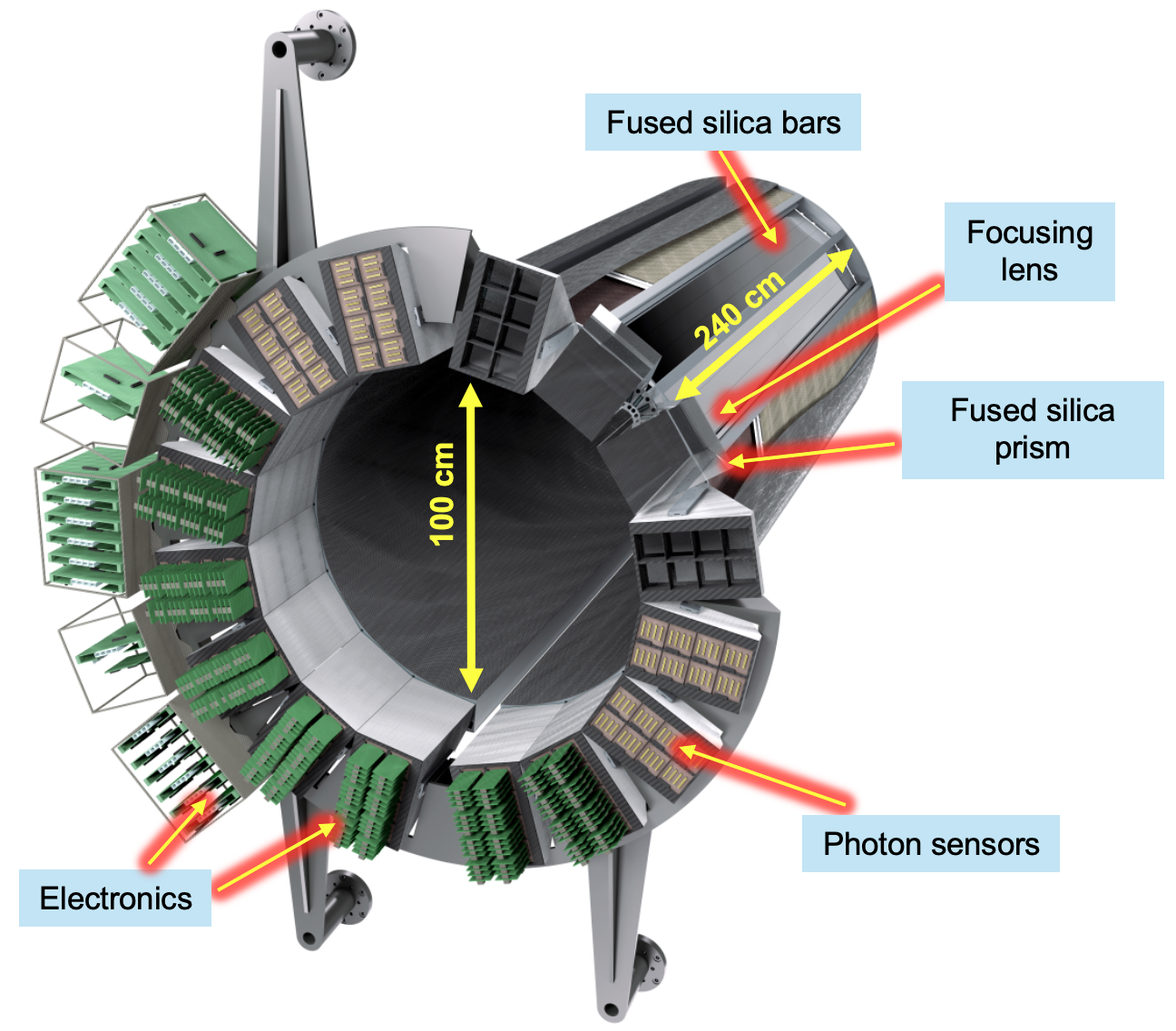}  
  \caption{Geant visualization of the PANDA Barrel DIRC geometry.}
  \label{Barrel DIRC}
\end{figure}

The PANDA\cite{PANDA},\cite{PANDAphysics} Barrel DIRC (Detection of Internally Reflected Cherenkov Light) detector \cite{tdr} features a modular design comprising 16 optically isolated sectors arranged in a barrel configuration with a radius of approximately 50 cm around the beam line (see Fig. \ref{Barrel DIRC}). Each sector consists of two main components: a bar box and a readout box, which house the optical elements. The bar boxes are constructed from carbon-fiber-reinforced polymer (CFRP) material, while the readout boxes are fabricated from a lightweight, high-strength aluminum alloy using 3D printing. These low-Z materials were selected to minimize the impact on the electromagnetic calorimeter surrounding the Barrel DIRC.

Each bar box contains three radiator bars, each made up of two synthetic fused silica bars measuring 1200 mm in length with a cross-section of 17 $\times$ 53 mm$^2$, glued end-to-end. A flat mirror is attached to the forward end of each radiator bar to reflect Cherenkov photons toward the readout end. At the end of each bar, a three-layer spherical lens focuses the Cherenkov photons onto an array of eight Microchannel-Plate Photomultiplier Tubes (MCP-PMTs), which are mounted on the rear surface of a fused silica expansion volume (EV) of 30~cm depth housed in a readout box.

The detector employs fast FPGA-based electronics capable of detecting single photons with a timing precision of approximately 100~ps. The entire system is supported by a lightweight structure, ensuring stability and precision while maintaining minimal material interference with other detector components.
\section{Lens focusing and fast photon timing}
\begin{figure}[h]
  \centering
  \includegraphics[width=.51\columnwidth]{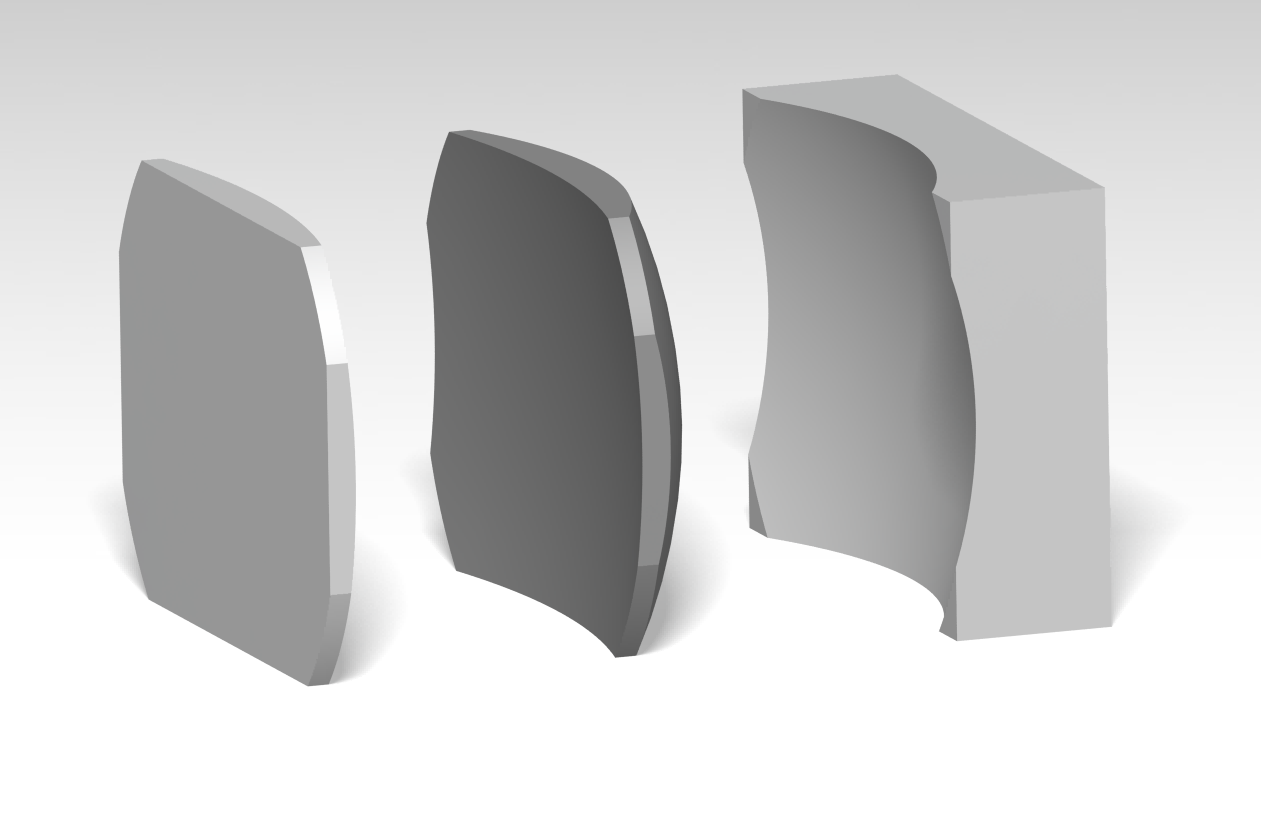}   
  \includegraphics[width=.33\columnwidth]{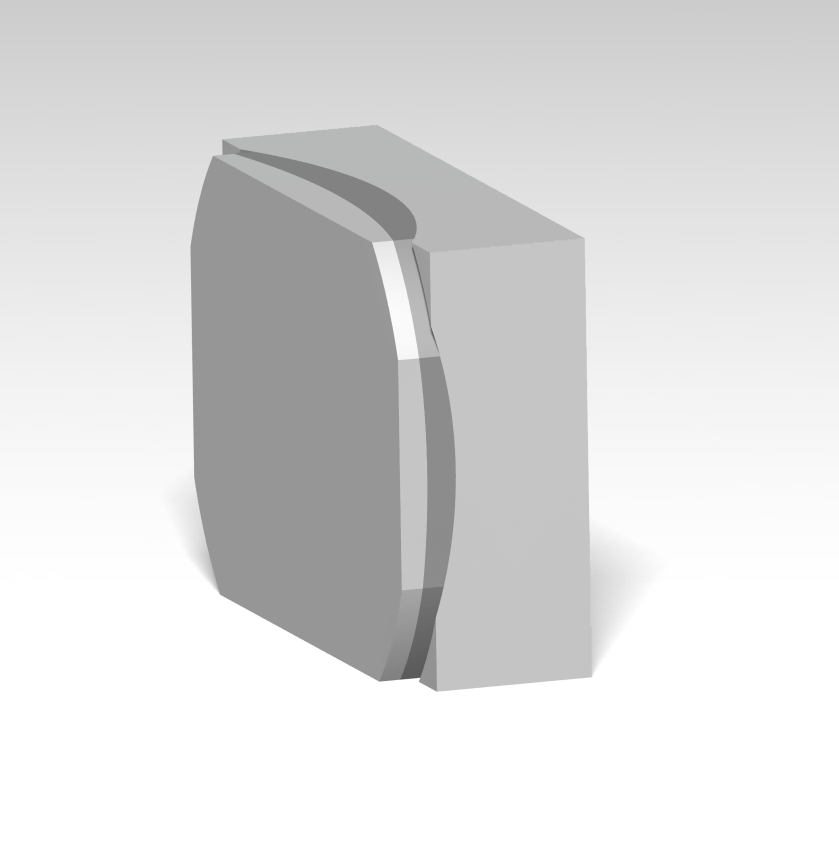}   
  \includegraphics[width=.1345\columnwidth]{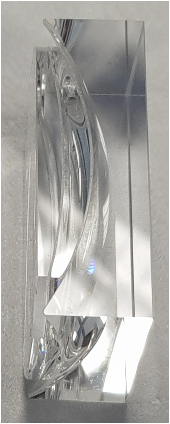}  
  \caption{The three layer spherical lens of the PNDA Barrel DIRC. The CAD-schematics on the left and in the middle show the exploded, respectively the compound view of the lens with the lanthanum crown glass (in black) between two synthetic fused silica pieces (in grey). The photo on the right shows a prototype lens.}
  \label{optics}
\end{figure}

The first DIRC detector design with lens focusing has been realized with the PANDA Barrel DIRC using an innovative three-layer spherical lens \cite{lens}. This lens design enables the formation of sharp ring images on a nearly flat focal plane at the end of the EV, significantly improving the detector performance. The focusing capability allows for the use of wider radiator bars, reducing the number of surfaces that need polishing. A key feature of this design is the incorporation of a high-refractive-index material, lanthanum crown glass, sandwiched between synthetic fused silica layers (see Fig. \ref{optics} left and middle). This eliminates air gaps between the lens and the radiator, respectively the lens and the EV (see Fig. \ref{optics} right), thereby minimizing photon loss that would occur with a single-lens system.
\vspace{0ex}

The photon sensors used are 8x8 pixel MCP-PMTs with a 6 mm pitch size and a position resolution of 2 mm. These sensors are optimized for fast and efficient single-photon detection within a 1~T magnetic field, achieving a timing precision of 100~ps. To meet the expected integrated anode charge of 5~C/cm$^2$ over the PANDA experiment's lifetime, these MCP-PMTs feature a lifetime-enhanced design achieved through atomic layer deposition coating on the MCPs.

\begin{figure}[h]
  \centering
  \includegraphics[width=0.9\columnwidth]{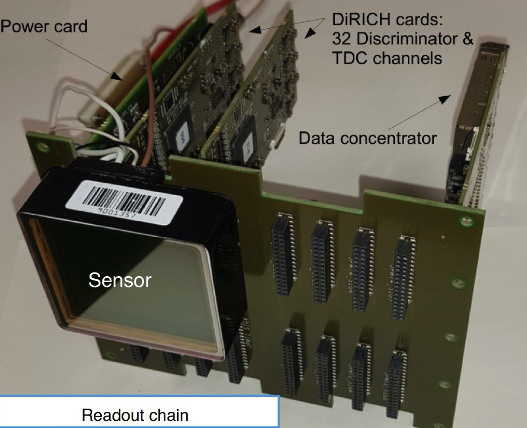}  
  \caption{DiRICH readout chain with MCP-PMT.}
  \label{DiRICH}
\end{figure}

The electronics system was developed in collaboration with the HADES/CBM RICH and includes the DiRICH readout module \cite{DiRICH1}, \cite{DiRICH2} (see Fig. \ref{DiRICH}). Each backplane connects four MCP-PMTs to two DiRICH cards, which provide both time and time over threshold information. The highly integrated design minimizes cabling and achieves an internal time precision of 10 ps (discriminator plus TDC). 
\section{Beam tests}
\begin{figure}[h]
  \centering
  \includegraphics[width=1.\columnwidth]{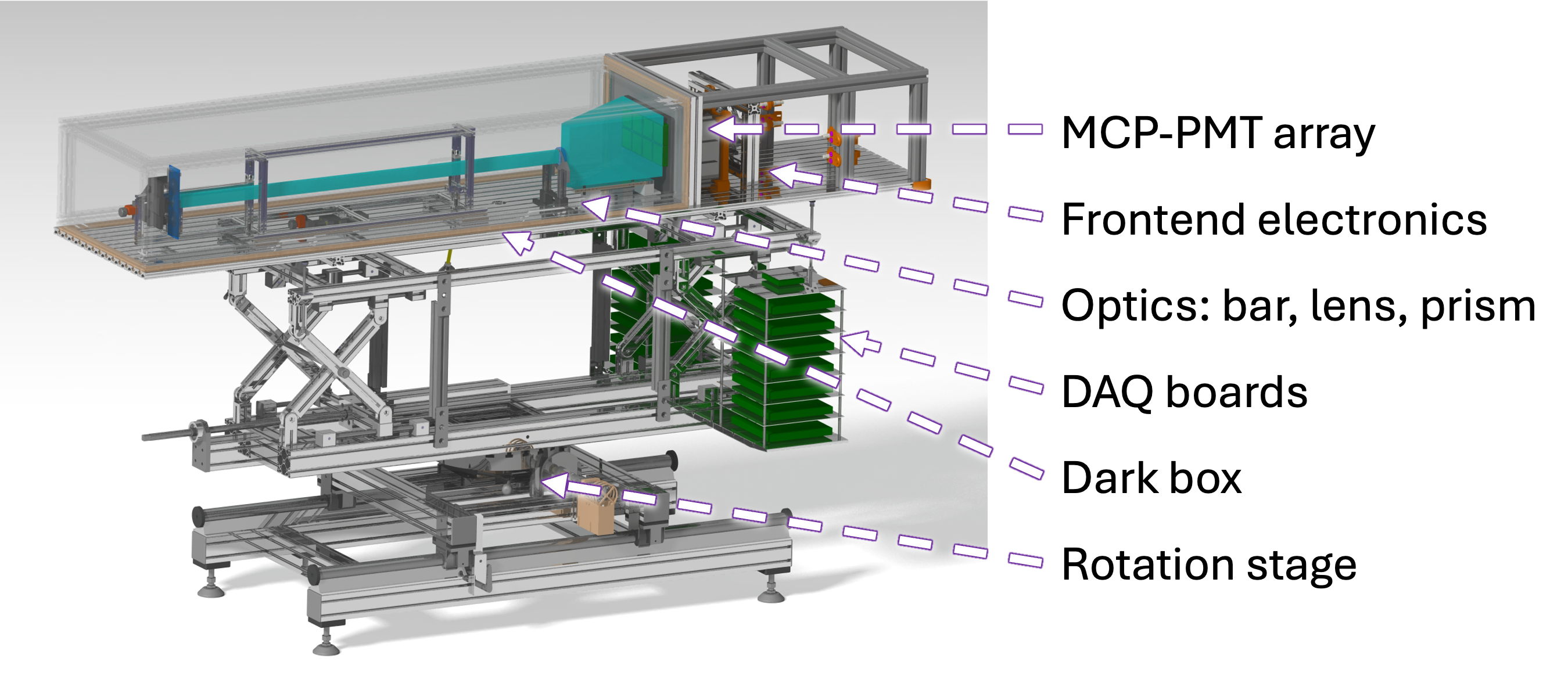}  
  \caption{CAD model of the prototype setup at CERN T9.}
  \label{Prototype}
\end{figure}

\begin{figure}[h]
  \centering
  \includegraphics[width=1.\columnwidth]{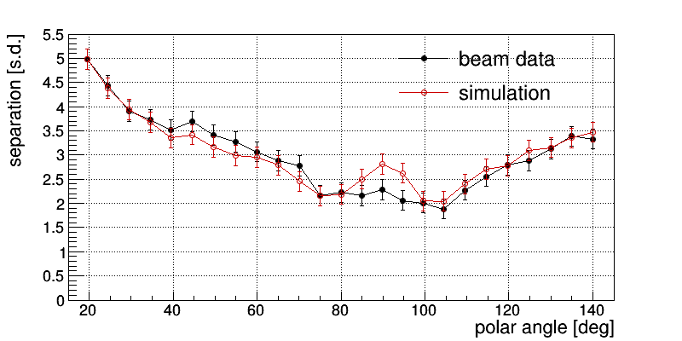}  
  \caption{$\pi$/p separation power vs. the beam polar angle at 7 GeV/c.}
  \label{Separation}
\end{figure}

The design of the Barrel DIRC was successfully validated in particle beam tests at CERN. In the hadron-rich beam of the T9/PS area, a prototype featuring a radiator bar, a three-layer spherical lens, a prism-shaped EV, and an array of eight PHOTONIS MCP-PMTs was tested (Fig. \ref{Prototype}). The setup demonstrated excellent performance, with the key performance parameters measured over the polar angle range from 22$^\circ$ to 140$^\circ$ at various beam momenta.
\vspace{0ex}
The results shown in Fig. \ref{Separation} show that for $\pi$/p separation at 7 GeV/c, the measured separation power was in excellent agreement with simulations. With 10 to 60 detected photons and a Cherenkov angular resolution of 6.9 mrad per photon, a $\pi$/p separation of 5~s.d. was achieved at a polar angle of 20$^\circ$. This performance corresponds to the separation power for $\pi$/$K$ identification at 3.5 GeV/c, confirming that the PANDA Barrel DIRC meets or exceeds the particle identification requirements set for the PANDA experiment.
\section{Component series production and quality assurance}
\begin{figure}[h]
  \centering
\includegraphics[width=1.\columnwidth]{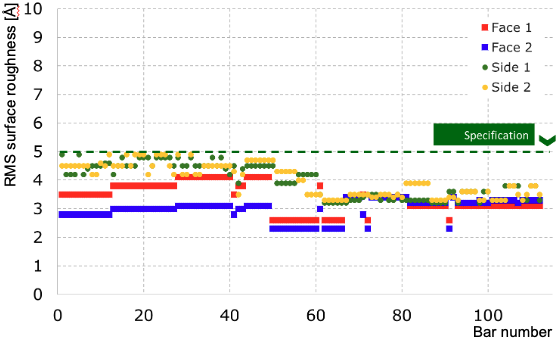}
  \caption{Nikon QA measurements using the example of surface roughness. All the delivered radiator bars meet this specification.}
  \label{Nikon-Measurements}
\end{figure}

The Technical Design Report \cite{tdr} was completed in 2017.
The series production of the Barrel DIRC radiator bars began in September 2019, with the fused silica bars fabricated by the Nikon Corporation, Japan. A total number of 96 highly polished radiator bars, each with a surface roughness of $\leq$~5$\AA$ and a deviation from parallelism and squareness of  
$\leq$~0.5~mrad, was required for the Barrel DIRC. Including spares, 112 bars were ordered, that Nikon delivered ahead of schedule between March 2020 and February 2021. The bars underwent rigorous quality assurance processes to ensure compliance with all specified properties, including surface roughness (see Fig. \ref{Nikon-Measurements}), squareness, parallelism, and total thickness variation. These measurements were conducted in-house by Nikon and confirmed that all bars met the fabrication specifications.

Detailed quality assurance measurements on the radiator bars in the optical lab at GSI focus on determining internal surface roughness, a parameter not measured by the radiator producers. This aspect is particularly significant because surface treatments, such as polishing, can potentially cause subsurface damage, which in turn may reduce the photon transport efficiency.
\begin{figure}[h]
  \centering
  \includegraphics[width=1.\columnwidth]{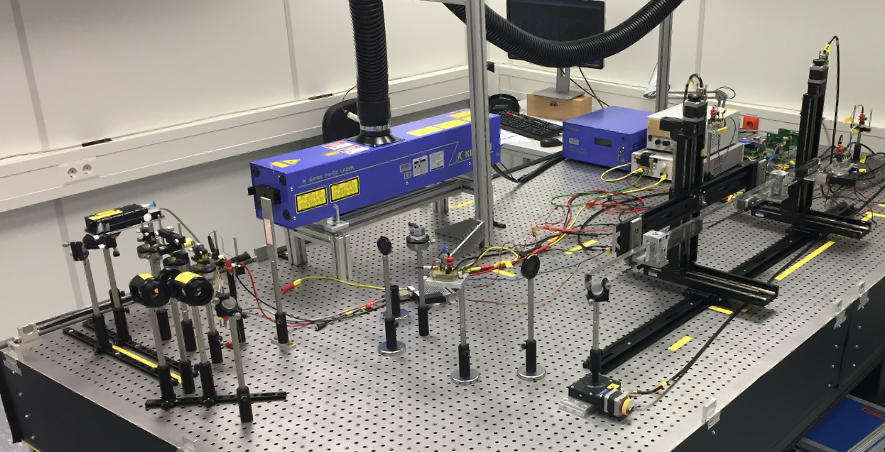}  
  \caption{Laser setup for the determination of the internal surface roughness of the radiator bars.}
  \label{Laser-Set}
\end{figure}
\vspace{0ex}
The method used for the measurements involves a laser setup (Fig. \ref{Laser-Set}) equipped with lasers of various wavelengths in the near UV (266 nm, 325 nm) and visible range (405 nm, 442 nm, 532 nm, and 635 nm), and six remote-controlled motors.

\begin{figure}[h]
 \centering
 \includegraphics[width=1.0\columnwidth]{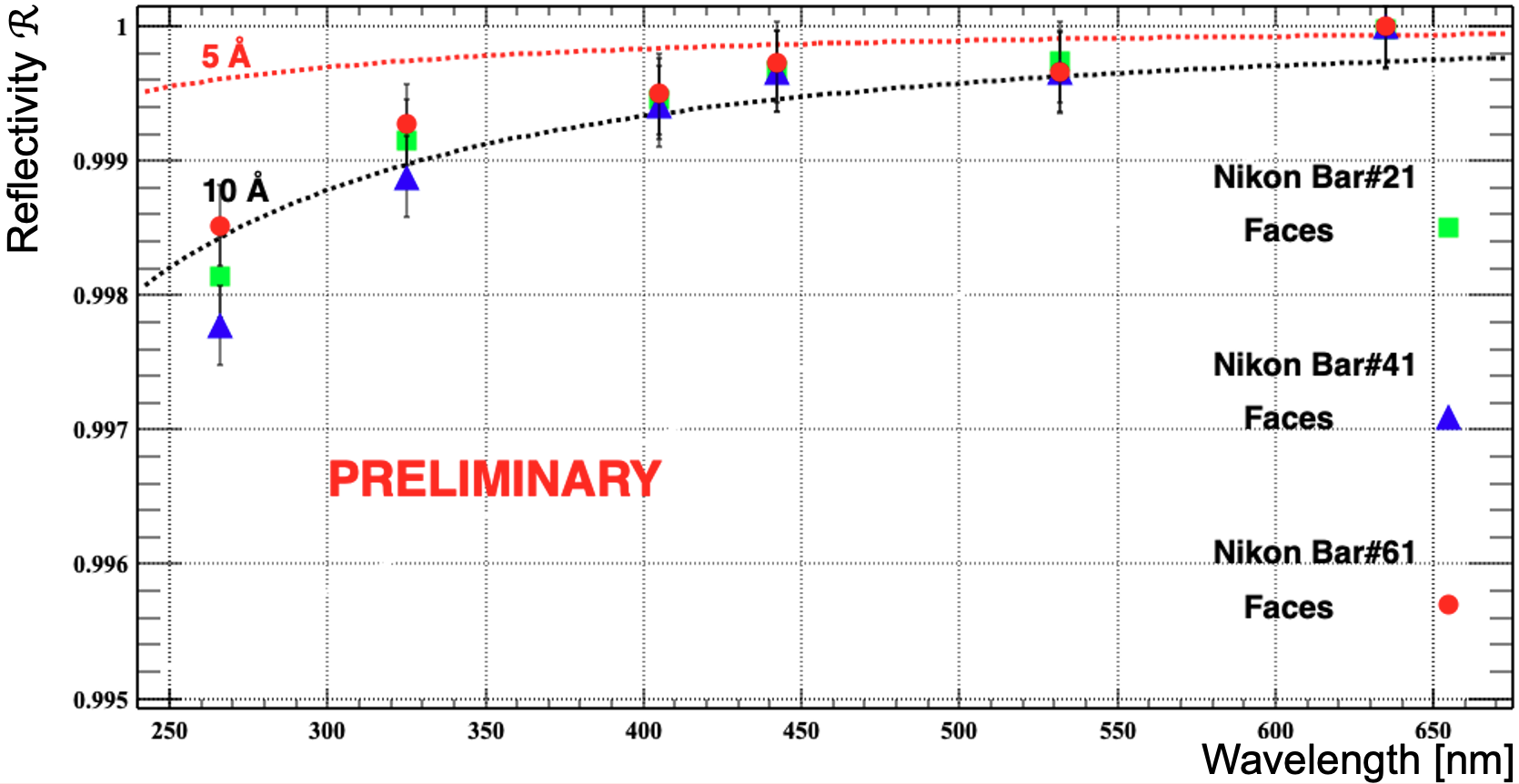}
 \caption{Preliminary results of the coefficient of total internal reflection vs. the laser wavelength for several of the Nikon bars.}
 \label{MeasurementsGSI}
 \end{figure}

Horizontally polarized laser beams enter the radiator bars at Brewster angle, ensuring loss-free transmission. Comparing the measured intensity of the split beam from two diodes, fluctuations in the original laser beam intensity are suppressed.
Using Scalar Scattering Theory \cite{STT}, the internal surface roughness is deduced from the wavelength dependence of the reflectivity measurements. The results (see Fig. \ref{MeasurementsGSI}) show no significant subsurface damage caused by the surface polishing method used by Nikon. 

The series production of the MCP-PMTs by PHOTONIS Netherlands BV started in 2022. The $2\times2~inch^2$ XP85112-S-BA MCP-PMTs have MCPs with 10~$\mu m$ pore diameter.  The first batch of the 155 ordered sensors was delivered in May 2022, and as of now, the production is approximately 50$\%$ complete. Quality assurance tests for these MCP-PMTs have been conducted at the University of Erlangen, focusing on the key parameters gain, quantum efficiency, dark count rate, cross talk, afterpulsing, time resolution, and rate stability, collection efficiency, behavior in magnetic fields of up to 1 Tesla and lifetime \cite{sensors}, \cite{sensorsProc}.

\begin{figure}[h]
  \centering
\includegraphics[width=1.\columnwidth]{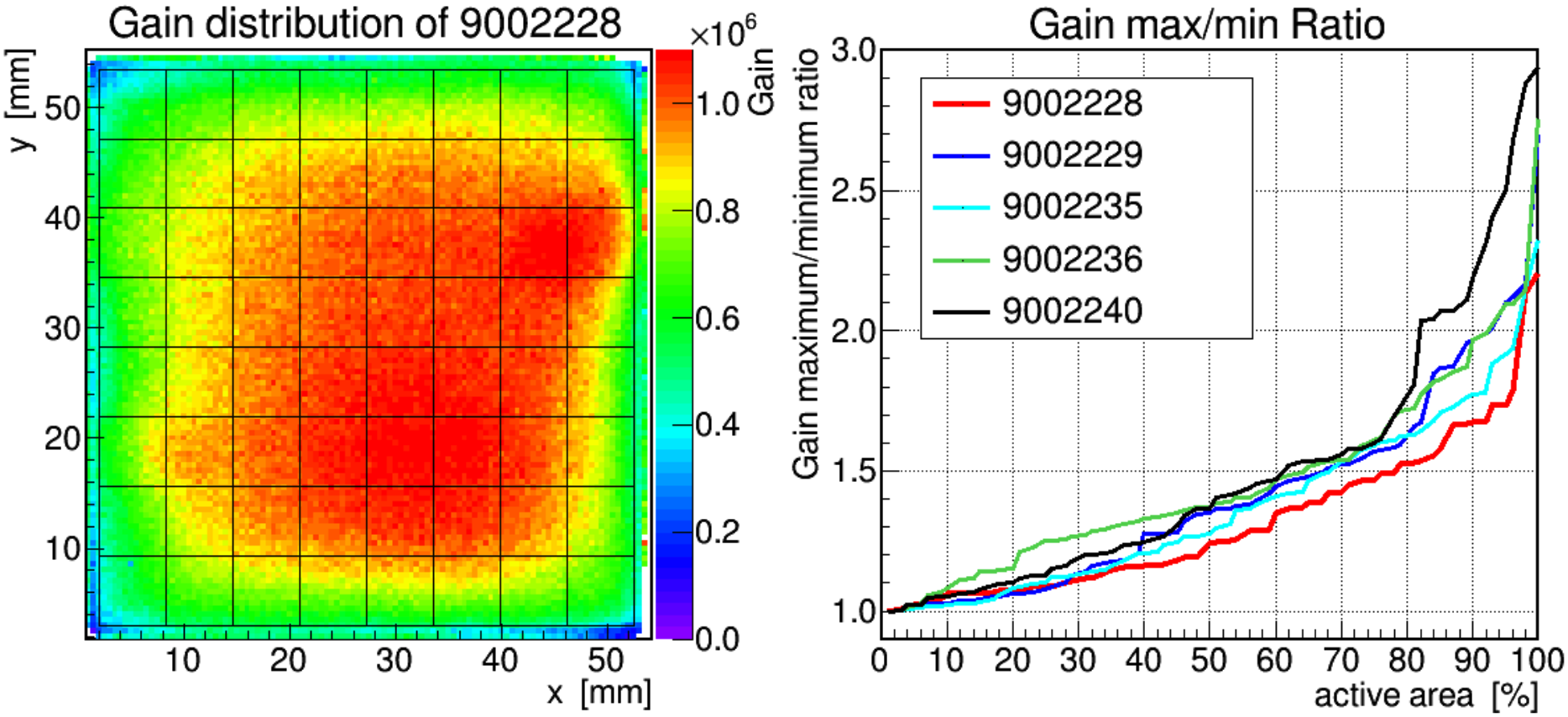}
  \caption{Measurements of the gain distribution over the MCP-PMT (left) and gain ratio for several tubes over the active area fraction (right). }
  \label{Gain}
\end{figure}

One example of the measurements on the tubes is a detailed gain scan performed with fine steps ($\leq$~0.5~mm) using single photons at a laser wavelength of 372~nm with a frequency of 50~kHz and a gain of $10^6$ (Fig. \ref{Gain} left). These tests revealed lower gains at the edges and corners of the tubes. However, the specification requiring a maximum-to-minimum gain ratio of $\leq$~3 across 90$\%$  of the active area was achieved for the tested devices (Fig. \ref{Gain} right). 
\section{Material tests and assembly procedure}
\begin{figure}[h]
  \centering
\includegraphics[width=0.495\columnwidth]{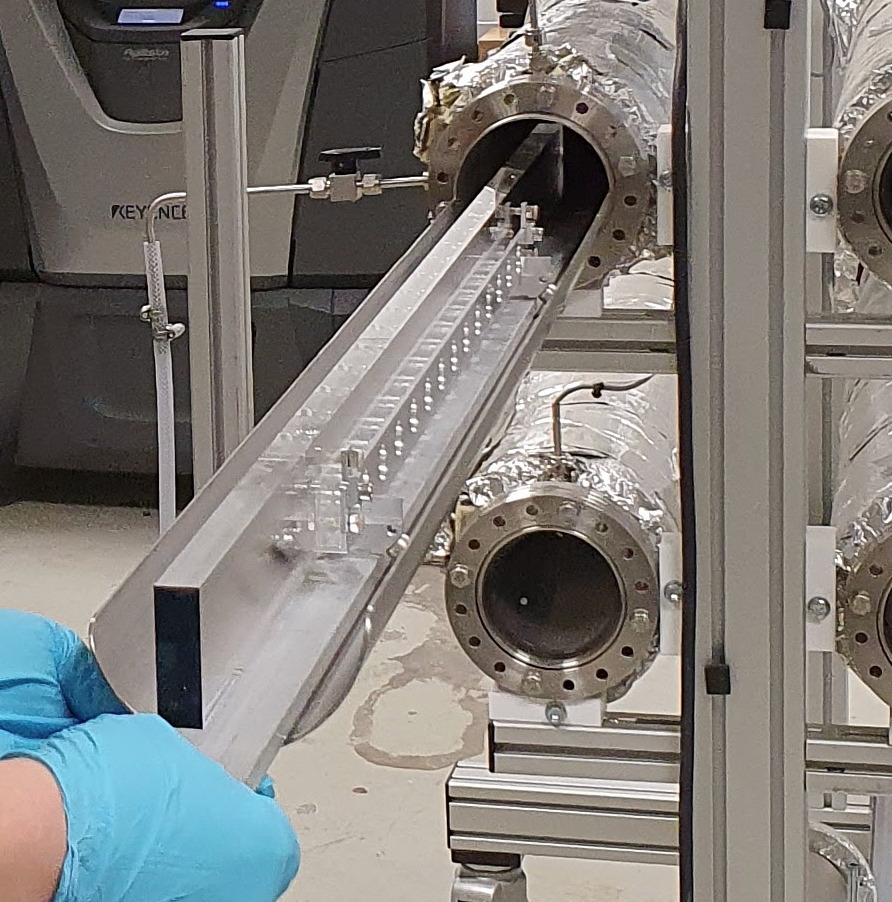}
\includegraphics[width=0.49\columnwidth]{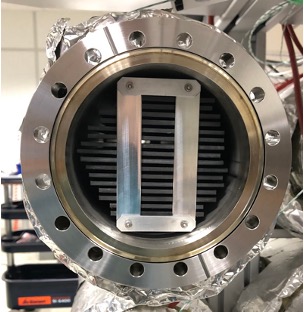}
  \caption{Material screening setup. Insertion of the radiator bar into the tube (left) and the stack of CFRP material in the connected vessel (right). }
  \label{Material screening}
\end{figure}

\begin{figure}[h]
  \centering
\includegraphics[width=1.\columnwidth]{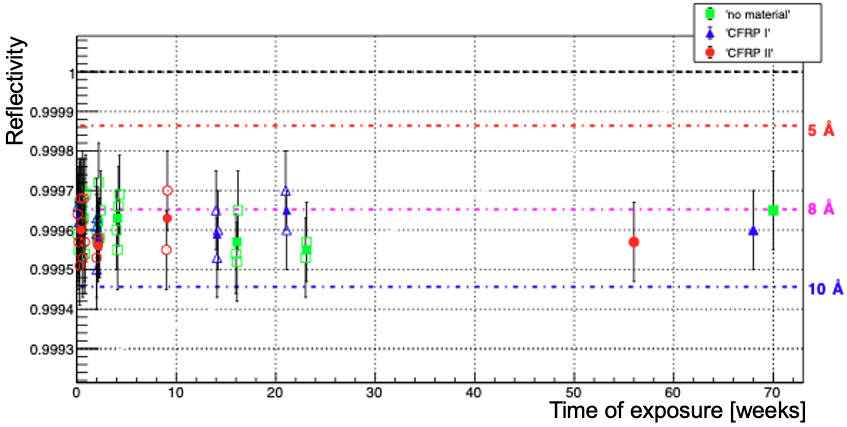}
  \caption{Coefficient of total internal reflection versus time of exposure for a reference bar and two bars exposed to different types of CFRP samples.}
  \label{LongtermMeasurementsGSI}
\end{figure}

Previous DIRC counters used aluminum panels to build the bar boxes. The PANDA Barrel DIRC as the first will profit from the advantageous properties of the CFRP, i.e. light weight and stiffness. It is essential to assess the potential long-term effects of outgassing from this material, as well as from other components within the bar boxes, on the surfaces of the DIRC bars. Any contamination could negatively impact the photon transport efficiency. To investigate this, several bars were placed inside stainless steel 
tubes (Fig. \ref{Material screening} left) each connected to a vessel with a large sample of one of the various CFRP candidate materials (Fig. \ref{Material screening} right). Many layers of thin CFRP sheets were stacked to maximize the total surface area for this test. Nitrogen gas continuously flowed across the materials and the bar surfaces for up to 18 months.
The reflection coefficient of these bars was measured before the exposure and later periodically using the laser setup with the 442 nm HeCd laser. The results, shown in Fig. \ref{LongtermMeasurementsGSI}, indicate no significant impact on the reflection coefficient of bars exposed to two different CFRP materials, compared to a reference bar that was kept in a clean nitrogen atmosphere. This demonstrates that the tested CFRP materials do not pose a risk of pollution affecting the photon transport efficiency under the given conditions.

\begin{figure}[h]
  \centering
\includegraphics[width=1.\columnwidth]{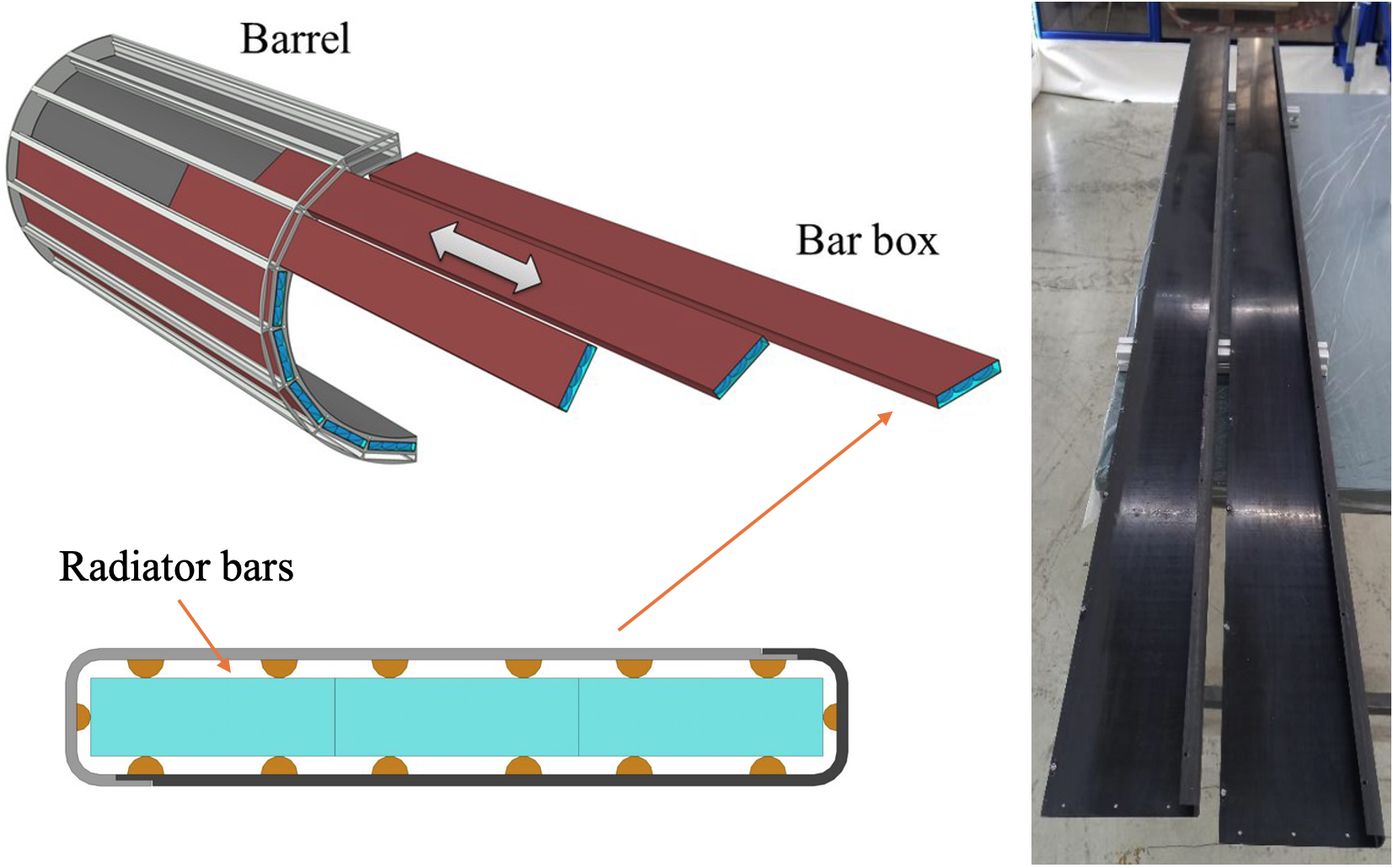}
  \caption{The bar box houses 3 radiator bars (left, down) and is stacked into the holding structure (left, up). Two L-shaped CFRP sheets form a bar box (right).}
  \label{CFRP-stiffness-test}
\end{figure}
\vspace{0ex}
\begin{figure}[h]
  \centering
\includegraphics[width=0.49\columnwidth]{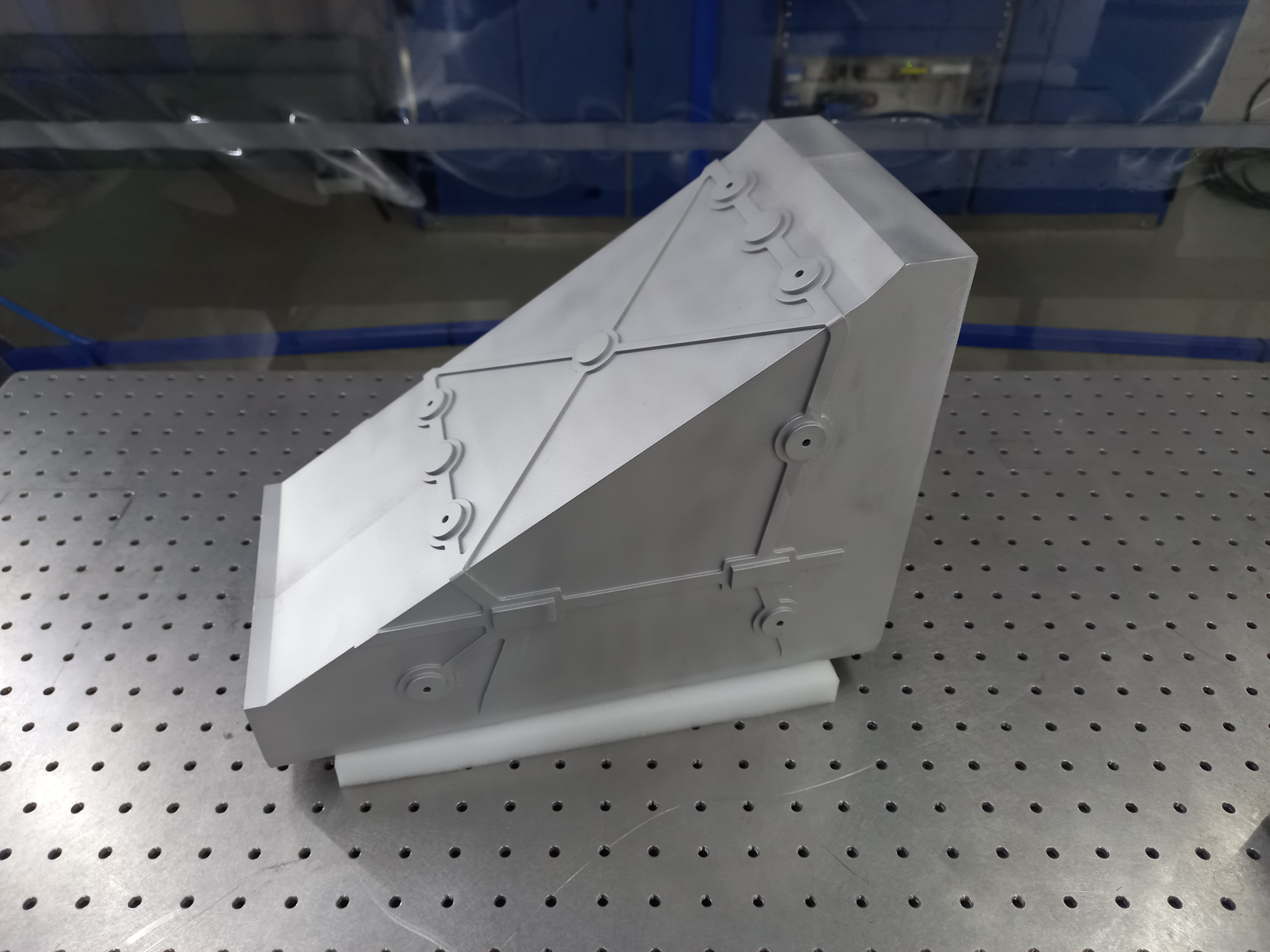}
\includegraphics[width=0.453\columnwidth]{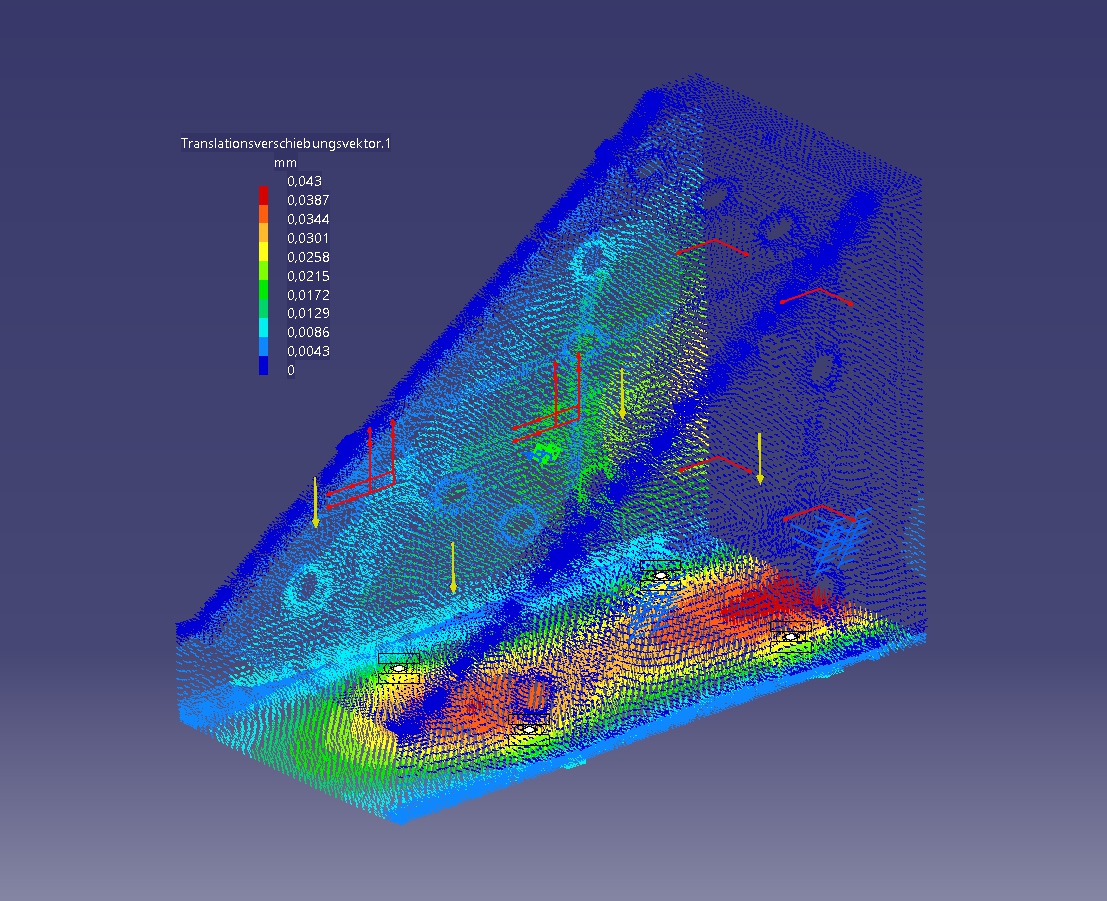}
  \caption{Readout box from 3D-printed aluminum alloy (left) and FEA studies (right). }
  \label{Prism-FEcalc-test}
\end{figure} 

The assembly procedure for a sector of the PANDA Barrel DIRC detector will be tested using prototypes of containers designed for the DIRC bars (Fig. \ref{CFRP-stiffness-test} right) and the prism EV (Fig. \ref{Prism-FEcalc-test} left), both produced by industry. 
The two main elements of the bar box are the L-shaped CFRP panels, which are closed by an endcap, containing the mirror assembly, and a window at the readout end.  The modular design allows that the bar boxes can be inserted into a holding structure one by 
one (Fig. \ref{CFRP-stiffness-test} left).
The bar box prototype will be loaded with aluminum dummy bars that match the weight and size of the actual fused silica radiator bars to test the stiffness of the CFRP construction. The results from these tests will be compared with finite element analysis (FEA) calculations to validate the stiffness of the mechanical design. As an example of such FEA calculations, stress distributions in a fully equipped 3D-printed lightweight aluminum alloy readout box have been analyzed (Fig. \ref{Prism-FEcalc-test} right).

\section{Summary}
The PANDA Barrel DIRC, featuring the first focusing lens design of a DIRC detector, fast photon timing and lifetime-enhanced MCP-PMTs paired with a DiRICH readout, has been successfully evaluated in particle beam tests. The production of the bars was successfully completed and the series production of the MCP-PMTs is underway. The QA shows that these components meet the PANDA Barrel DIRC requirements. For the bars no significant subsurface damage was detected.
Prototypes of the detector mechanics, produced by industry, are now being prepared for assembly testing. One sector of the modular Barrel DIRC is planned to be set up to validate the assembly procedure comprehensively.
\\
This work was supported by BMBF and HFHF. 

\end{document}